\documentclass[fleqn,11pt]{wlscirep}
\usepackage[utf8]{inputenc}
\usepackage[T1]{fontenc}
\usepackage{subfigure}
\usepackage{enumitem}
\setlist{leftmargin=3.5mm}
\usepackage{bbm}
\usepackage{multirow}
\usepackage{gensymb}
\title{High-precision Density Mapping of Marine Debris and Floating Plastics via Satellite Imagery}

\author[1,2]{Henry Booth}
\author[1]{Wanli Ma}
\author[1,*]{Oktay Karakus}
\affil[1]{Cardiff University, School of Computer Science and Informatics, Abacws, Cardiff, CF24 4AG, UK}
\affil[2]{Met Office, FitzRoy Road, Exeter, Devon, EX1 3PB, UK}

\affil[*]{Corresponding Author - karakuso@cardiff.ac.uk}

\begin{abstract}
Combining multi-spectral satellite data and machine learning has been suggested as an effective method for monitoring plastic pollutants in the ocean environment. Recent studies have made theoretical progress regarding the identification of marine debris and suspected plastic (MD\&SP) via machine learning. However, no study has fully explored the application of these methods for mapping and monitoring marine-debris density. As such, this paper comprised of three main components: (1) the development and validation of a supervised machine learning marine debris detection model, (2) embedding it into an automated tool named \textit{MAP-Mapper} for mapping MD\&SP density, and finally (3) an evaluation of the whole system for out-of-distribution (OOD) test locations. 
The developed MAP-Mapper architectures provide users choices to reach high-precision (\textit{abbv.} -HP) or optimum precision-recall (\textit{abbv.} -Opt) values in terms of the training/test data set. Our MAP-Mapper-HP model greatly increased the precision of MD\&SP detection to 95\%, whilst MAP-Mapper-Opt reaches precision-recall pair of 87\%-88\%. In order to efficiently quantify the density mapping findings in OOD test locations, we propose the \textit{Marine Debris Mapping} (MDM) index, which combines the average probability of a pixel belonging to the marine debris class and the number of detections within the given time interval. 
High MDM findings of the proposed approach have been found to be aligned with the existing marine debris and plastics pollution areas, and these are presented with existing evidence referring to the literature and field studies.
\end{abstract}

\begin{document}

\flushbottom
\maketitle
%
%
\thispagestyle{empty}


\section*{Introduction}
Current estimates indicate that there are now millions of tons of plastic floating in the world's oceans, with millions more entering each year\cite{lebreton2018evidence}. Since plastics have an extremely slow rate of decomposition, marine plastic is rapidly accumulating. 
Whilst micro-plastics (0.05-0.5cm) and mesoplastics (0.5-5cm) are by far the most numerous, macro (5-50cm) and mega-plastics ($>$50cm) are thought to make up the majority of the total weight of ocean plastic\cite{lebreton2018evidence}. Therefore, effective monitoring and mapping of larger plastic objects are needed to answer key scientific questions regarding the sources, distribution, and transportation of plastic in the ocean environment. These insights could help advise preventative measures and clean-up operations, improving their efficacy\cite{lebreton2019global}. 
In recent years, machine learning techniques have been successfully applied for classification in various vision-related areas such as healthcare, text analytics, cybersecurity, and geo-sensing problems such as road extraction and land cover mapping\cite{sarker2021deep,zhang2018road,ma2022amm}. Furthermore, advances in this field have led to models that can outperform human experts in some tasks \cite{buetti2019deep,zhou2021ensembled}. Therefore, machine learning algorithms can remove the need for manual classification, whilst maintaining and sometimes improving performance. This creates opportunities for the automatic classification of marine plastic. Therefore, combining satellite data with machine learning algorithms have been suggested as a method to automatically survey large areas for plastic pollution\cite{martinez2019measuring,michel2020rapid}. 

In an initial study, Aoyama\cite{aoyama2016extraction} developed a method for identifying marine plastic by using two-dimensional scatter diagrams of satellite spectral bands. Pixels that had large spectral differences from the surrounding ocean were suspected to be plastic. The method was then validated using a known target of a fixed fishing net with buoys. Furthermore, Topouzelis et. al. \cite{topouzelis2019detection} created and deployed three man-made plastic targets off the coast of the Island of Lesvos for Plastic Litter Project (PLP). The objective of this work was to assess the spectral signatures of plastic targets in Sentinel-2 satellite imagery
. Researchers used high-resolution aerial imagery to calculate the percentage of plastic coverage for each pixel in the satellite image. This enabled investigation into how the spectral signature of each Sentinel-2 pixel changed as plastic coverage decreased. It was suggested that these findings would help assess which wavelengths offer the best opportunity to differentiate 
sea foam, white caps, and surface-reflected glint from marine debris. 

Moy et al.\cite{moy2018mapping} used aerial surveys and manual inspection of images to map the quantity and location of macro-plastics on the coastline of the eight main Hawaiian islands. They found that windward shorelines had the highest density of plastic and produced a map of the coastline to visualise this. This study demonstrates the ability to map marine plastic on coastlines using aerial imagery. However, due to manual inspection, evaluating the entire coastline in this way is very laborious. 
Biermann et al.\cite{biermann2020finding} demonstrated that it was possible to train a machine learning algorithm to differentiate between plastic and other types of marine debris. The authors proposed a novel parameter, which is the floating debris index (FDI), in order to analyse sub-pixel interactions of macroplastics with the sea surface to increase the chance of detection of floating patches. 
Their model utilised Sentinel-2 data and was trained on various marine debris targets manually extracted with the help of FDI and normalised vegetation difference index (NDVI) from Scotland, British Columbia, Barbados and Durban whilst the plastic targets developed by Topouzelis et al.\cite{topouzelis2019detection} used for validation purposes. This ground-truth data ensured validity and maximum accuracy of 86\% was reported for the classification of plastics among seaweed, seawater, timber, and plume. 

Kikaki et al.\cite{kikaki2020remotely} investigated the plastic pollution problem in the Bay Islands of Honduras by using remote-sensing observations from 2014 to 2019. It was noted that detectable plastics generally follow linear patterns. Similarly to other plastic-mapping work, this was a labour-intensive task that lacks the automation required for global applicability and continual monitoring. In another recent work, 
Kikaki et al.\cite{kikaki2022marida} produce a Marine Debris Archive (MARIDA) which contains 1381 patches with 837,357 annotated pixels from 63 Sentinel-2 scenes acquired between 2015 and 2021. The patches are distributed over eleven countries. MARIDA contains 3339 (~0.4\%) Marine Debris pixels in total which were defined as "floating plastic and polymers, mixed anthropogenic debris". Of these plastic pixels, 1625 pixels were digitised and annotated with high confidence. 
This study also investigated the effectiveness of different machine-learning algorithms in classifying marine debris. Three variations of the random forest model were examined, as well as a U-net model where Random Forest models outperformed the U-net model. When discussing the MARIDA dataset and the models developed using it, marine debris will be used along with the term "suspected plastic" for the rest of this work. This is to increase clarity since plastic is thought to make up the vast majority of floating anthropogenic debris\cite{roman2016anthropogenic}. We also believe that this wary approach in defining the detected pollutants is also aligned with the cautious note on spectral interpretations published by Hu recently in \cite{hu2022remote}.


The primary aim of this work is to assess the feasibility of mapping marine debris and suspected plastic density on multi-spectral satellite imagery by using a machine learning algorithm. Whilst reaching the aforementioned aim, we (1) optimise a machine learning algorithm to enable precise detection of pollutants on the ocean surface, (2) develop an automated data pipeline that is capable of gathering, pre-processing and making predictions on satellite data, (3) use the data-pipeline to generate marine debris density maps for the test locations. 
To address the aforementioned aims, this work proposes a novel end-to-end automated system consisting of two main components: (A) a high-precision marine debris and suspected plastic detection machine learning algorithm (MAP-Mapper), and (B) a scientific tool/pipeline to facilitate marine-debris density mapping for any region of interest (ROI) in any given time frame. We evaluate the proposed detection algorithm with MARIDA data set whilst Topouzelis et. al.'s PLP data sets are used to validate the MAP-Mapper architectures. Six test locations including highly polluted areas such as Manila - Philippines, and Mumbai - India are used to test the proposed MAP-Mapper data pipeline and density mapper system. In order to efficiently quantify the density mapping findings, we proposed a pixel-level parameter - the Marine Debris Mapping (MDM) index, that combines the average probability of belonging to the marine debris class and the number of detections within the given time interval. High MDM findings of the MAP-Mapper algorithms have been found to be aligned with the existing marine debris and plastics pollution areas, and these are presented with existing evidence referring to the literature and field studies.   

\section*{Results}

\subsection*{Model Evaluation with MARIDA Data Set}
Baseline models provided by Kikaki et al.\cite{kikaki2022marida} used to test MARIDA data set had significant problems in terms of run-time and/or precision. In particular, these models in Kikaki et al.\cite{kikaki2022marida} regularly miss-classified marine water, natural organic material, sea foam, ships, and clouds as marine debris. Since these objects and ocean features are much more common than suspected plastic pixels, extremely high precision is obviously needed for automated monitoring of marine debris\cite{thushari2020plastic}. A lower precision is likely to result in significant numbers of false positives. Consequently, current models lack the precision required and are likely to produce inaccurate marine debris and suspected plastic density maps. 

To provide a solution to the aforementioned problem, we conducted a development procedure of multiple U-net-based machine learning models to produce an optimised model and enable more accurate and high-precision density maps. Kikaki et. al.'s \cite{kikaki2022marida} MARIDA dataset was used to train and evaluate the proposed models. Particularly, we developed two different MAP-Mapper models to provide (1) high precision (\textit{abbv.} -HP) and (2) theoretically optimum - in terms of precision-recall values - (\textit{abbv.} -Opt).
The comparison results and performance metrics for MARIDA data tests are given in Table \ref{tab:res1}.

\begin{table}[ht]
\centering
\begin{tabular}{l|p{2cm}p{2cm}p{1.7cm}|p{2cm}p{2cm}p{1.7cm}}
\toprule
& & \textbf{MD\&SP$^+$} & & & \textbf{Overall}& \\
\hline
\textbf{Model} & \textbf{mIoU} & \textbf{Precision} & $F_1$\textbf{-Score}& \textbf{mIoU} & \textbf{Precision} & $F_1$\textbf{-Score} \\
\hline
\textit{Kikaki et. al.}\cite{kikaki2022marida} \textit{U-net} & 0.33 & 0.35 & 0.49 & 0.66 & 1.00$^*$ & 0.74\\
\hline
\textit{Kikaki et. al.}\cite{kikaki2022marida} \textit{RF} & 0.67 & 0.79 & 0.83 & 0.70 & N/A & 0.81\\
\hline
\textit{MAP-Mapper-Opt} & \textbf{0.78} & 0.87 & \textbf{0.88} & \textbf{0.89} & 1.00$^*$ & \textbf{0.94}\\
\hline
\textit{MAP-Mapper-HP} & 0.60 & \textbf{0.95} & 0.75 & 0.80 & 1.00$^*$ & 0.88\\
\bottomrule
\end{tabular}
\caption{\label{tab:res1}Metrics for plastic detection. $^+$ MD\&SP refers to Marine Debris and Suspected Plastics. $^*$\small{Due to the large class imbalance the overall precision of all classification models was very close to, and thus rounded to 1, regardless of other metrics.}}
\end{table}

\begin{figure}[ht!]
\centering
\includegraphics[width=\linewidth]{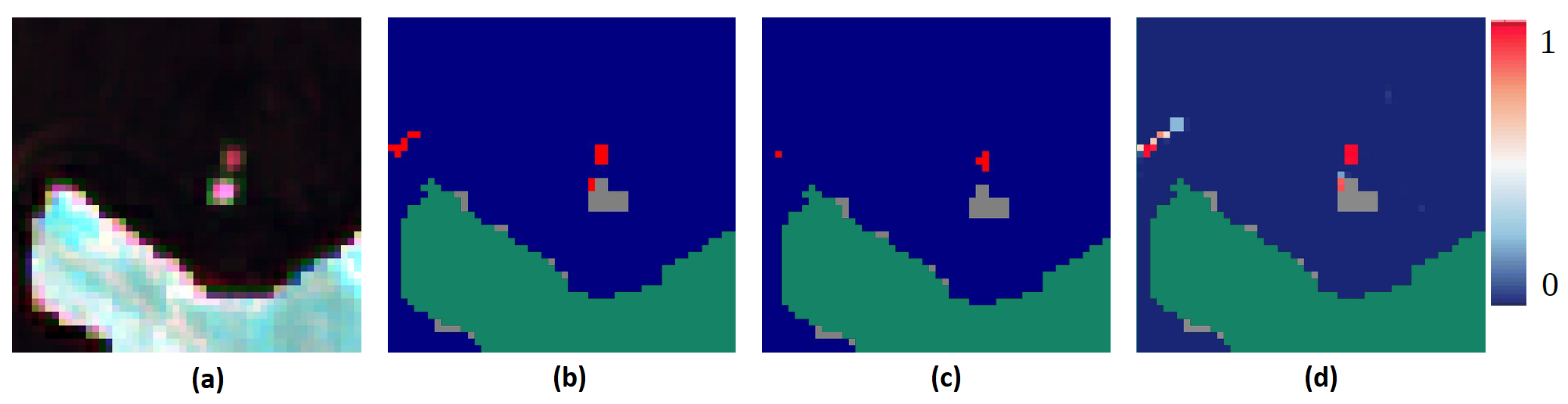}
\caption{Gulf of Lesvos suspected plastic detection. (a) Sentinel-2 false colour image, (b) MAP-Mapper-Opt, (c) MAP-Mapper-HP, (d) Probabilities. For the model in (b), 10 pixels of the HDPE and 6 of the wooden were classified correctly. 4 out of 6 of these pixels were then masked by F-mask. A number of false positives can be seen on the left of the figure, which appears to be the result of a ship or wake. For the model in (c), 4 HPDE pixels were classified as plastic, whilst no pixels from the wooden target were classified. 1 false positive was seen at a threshold of 0.99.} 
\label{fig:lesvos}\vspace{-0.35cm}
\end{figure}

\subsection*{Model Validation with PLP 2020 Data Set}
The Island of Lesvos, Greece has been the site of the Plastic Litter Project from 2018 to 2022. The experiment conducted by the University of Aegean’s Marine Remote Sensing Group in 2020 was selected for validation of the MAP-Mapper tools. The 2020 experiment was selected based on the certainty of the target position, variability in material, and size of the target. This meant that the model's ability to detect known plastic targets could be explored and its strengths and weaknesses could be assessed\cite{plp:2022}. For the PLP 2020 \cite{plp:2022}, two large targets were deployed, both approximately 28 meters in diameter (Figure \ref{fig:lesvos}-(a)). One comprised high-density polyethylene (HDPE) plastic mesh and the other was made of wood. 
At a threshold of 0.5, 10 pixels of the HDPE target and 6 of the wooden target were classified correctly. 4 out of 6 of these wooden pixels were then masked by F-mask. A number of false positives were observed which appear to be the result of a ship or wake. At a threshold of 0.99, 4 HPDE pixels were classified as suspected plastic, whilst no pixels from the wooden target were classified. 1 false positive was seen at a threshold of 0.99. F-mask appears to mask part of the natural wooden target but did not mask the plastic target.

\subsection*{MD\&SP Density Mapping}
This experiment case enables the assessment of MAP-Mapper model performances in different out-of-distribution regions and their global applicability. Marine debris density maps were produced from data spanning around a year for each test location. By following news about mass pollution events and also data from OpenOceans Global \cite{plasticmap_2022} and the OceanCleanUp river monitoring software \cite{cleanup_2022}, we decided on 6 test locations which are (1) the Gulf of Honduras, (2) Manila - Philippines, (3) Mumbai - India, (4) Hong Kong, (5) Cornish coastline, UK and (6) Chubut, Patagonia, Argentina. The first three test locations potentially accommodate high MD\&SP coverage locations and are selected to test proposed approaches in high-density scenarios whilst the remaining three are chosen to test the proposed approaches' capabilities in locations with no/low MD\&SP coverage. The details and data information of each test location are presented in Table \ref{tab:data}. 

MD\&SP density maps for each test location are shown in Figure \ref{fig:hond}. In each figure, the hexagonal width was set to 5km and for all plotted images each hexagonal area was fixed at approximately 22 km$^2$. Colour coding in each hexagon (vertical colour bar) shows 50\% left trimmed average MDM values considering each hexagon includes a multitude of pixels having 0 or very small MDM values. This enables us to remove pixels that do not contribute to the density analysis in each hexagon and to better quantify the novel MDM metric for the purposes of MD\&SP density mapping. For each test location, we also displayed the top 10 highest MDM values as scatter points. 

\begin{table}[t]
    \centering
    \begin{tabular}{p{4.5cm}p{2cm}p{2cm}p{4.5cm}p{2cm}}
    \toprule
Test Location &   Date Interval                & \# of Dates    & Location Coordinates               &    Area [km$^2$]  \\\toprule
Bay of Honduras            & 1 Jan 2022    & 25    & (16\degree11$'$56$''$N , 88\degree48$'$29$''$W)    & 5221.87 \\
& 13 Sep 2022 & & (15\degree40$'$38$''$N , 87\degree58$'$03$''$W) &\\\hline
Cornish coastline, UK    & 12 Jan 2022 & 30    & (50\degree31$'$13$''$N , 5\degree48$'$03$''$W)    & 6955.98 \\
& 26 Oct 2022 & & (49\degree56$'$39$''$N , 4\degree17$'$05$''$W) & \\\hline
Hong Kong                & 4 Jan 2022    & 37    & (22\degree21$'$11$''$N , 113\degree59$'$48$''$E)    & 320.96 \\
& 26 Oct 2022 & & (22\degree11$'$42$''$N, 114\degree10$'$26$''$E) & \\\hline
Manila, Philippines        & 7 Jan 2022    & 15    & (14\degree51$'$25$''$N , 120\degree39$'$13$''$E)    & 1605.43 \\
& 3 Nov 2022 & & (14\degree30$'$03$''$N, 121\degree01$'$51$''$E) & \\\hline
Chubut, Patagonia, Argentina            & 9 Jan 2022   & 44    & (45\degree13$'$07$''$S, 67\degree37$'$08$''$W) & 5846.64 \\
& 28 Oct 2022 & & (45\degree58$'$30$''$S, 66\degree43$'$13$''$W) & \\\hline
Mumbai, India            & 2 Jan 2022    & 21    & (19\degree33$'$43$''$N,
72\degree30$'$08$''$E)    & 5574.59 \\
& 3 Nov 2022 & & (18\degree39$'$31$''$N,
73\degree01$'$54$''$E) & \\\bottomrule
    \end{tabular}
     \caption{Details and data information of each test location evaluated for MD\&SP density mapping. }
     \label{tab:data}
\end{table}

\subsubsection*{Gulf of Honduras}
The Gulf of Honduras is an important inlet in the Caribbean Sea including coastal areas of Honduras, Guatemala and Belize. Even though it is seen as one of the important spots of a diverse and unique ecosystem due to the effects of strong ocean currents, the Gulf of Honduras is also one of the well-known plastic hotspots that have been the research site of numerous studies investigating marine-plastic pollution.

The MAP-Mapper density mapping results highlight three important regions in this test location. 
The first area is in the Bay of Amatique. Most of the detections appear to be contained within a relatively small area (4 hexagons in total). This location was identified as the coasts of Macho Creek, Puerto Barios, and Bahia La Graciosa. This area contains all the highest MDM-valued locations and has an average MDM of around 2.45 with maximum MDM values of around 7.00 (1 in Figure \ref{fig:hond}). 
The second region (2 in Figure \ref{fig:hond}) is off the coast of Punta Gorda. There is one hexagon with an average MDM value of 1.52.
The third area of extremely high MD\&SP density is noted off the coasts of Omoa, near the Motagua river mouth, which is located southwest of Omoa. This area has two highly dense hexagons with average MDM values of  around 1.00 (3 in Figure \ref{fig:hond}).

\subsubsection*{Manila - Phillipines}
Manila is the capital city of the Philippines sitting in the metropolitan area of Metro Manila which is the 5$^{th}$ populous metropolitan area in the world. Manila has also a negative reputation for having a pollution problem and is another important globally acknowledged MD\&SP hotspot via accommodating three out of five most pollutant rivers in the world \cite{plasticmap_2022}. 

Map-Mapper density mapping results are 
highlighted under three regions as given below.
The first region is Manila centre and its north coastlines. In this area, 6 hexagons ($\approx$130km$^2$) are having average MDM values higher than 1.00 (with the highest 2.87). All 10 highest pixel-level MDM values recorded within this area values of which are changing between 9.00 and 17.00 (5 in Figure \ref{fig:hond}). 
Cavite City having the highest average MDM value of 3.55 includes 3 hexagons higher than 1.00 MDM (6 in Figure \ref{fig:hond}). 
Compared to the other two regions highlighted above, the third region has less MD\&SP coverage however worth highlighting here. There are 2 densely polluted hexagons off the coasts of San Pascual, and 1 off the coast of Santa Cruz with average MDM values around 0.90 (4 in Figure \ref{fig:hond}). 

\subsubsection*{Mumbai - India}
Mumbai is the capital city of Maharashtra state of India and named as the 8$^{th}$ highest-populous city in the world. 
Mumbai
is known as highly polluted in terms of its seashores and inland waters especially due to having the world's 3rd most polluted river of Ulhas. 

The MAP-Mapper density mapping findings can be summarised under three regions:
The most MD\&SP problematic area of Mumbai is the south coasts of the main Mumbai Island coasts of Arabian Sea near Mahim Bay, Mumbai Harbour of Thane Creek, and Back Bay. This area has 4 highly densely polluted hexagons valued over 1.00 with the highest of 2.40 (7 in Figure \ref{fig:hond}). 
The second area is inside the city consisting of Navi Mumbai coasts, Panvel creek, and Thane Creek with three high average MDM hexagons with a maximum of 2.05 (8 in Figure \ref{fig:hond}). 
The mouth of Thane creek opening to the Arabian Sea off the coasts of Uran at the south of Mumbai includes two high MDM hexagons both of which are higher than 2.00 MDM with a highest of 2.77 (9 in Figure \ref{fig:hond}). 



\subsubsection*{Cornish coastline, UK \& Hong Kong \& Chubut, Argentina}
The Cornish coastline is located in the Southwest of England, also known as the Cornwall coastline. This area is one of the most tourist-visiting regions in the UK and is known for its outstanding natural beauty. 
A recent study has indicated that the Cornish coastline is the most plastic-polluted coastline in the UK\cite{nelms2020investigating}. However, the severity of the problem is far less than in the aforementioned test locations. 
Hong Kong is a special administrative region in South China, that can be listed as one of the most densely populated regions in the world. Despite its relatively smaller area of land, Hong Kong is one of the most important regions in the world in terms of import/export traffic. Chubut province in South Argentina, located in the Andes mountains to the west and the Atlantic Ocean to the east, is one of the important wildlife tourist places in the world especially via accommodating one of the globally largest Magellanic Penguin breeding areas. 

\begin{figure}[ht!]
\centering
\includegraphics[width=.99\linewidth]{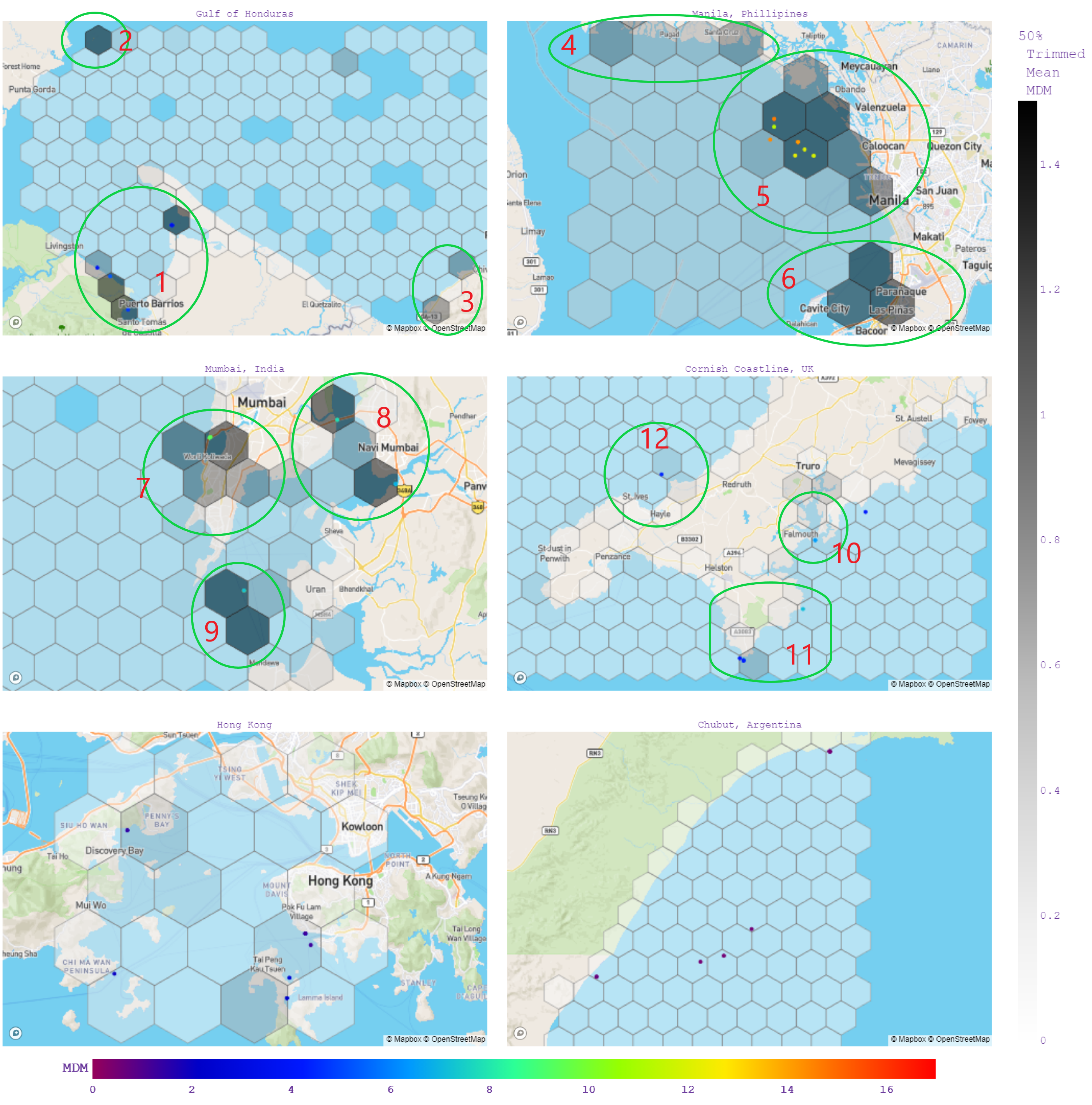}
\caption{Test location MD\&SP density map
s
. The colour bar on the right corresponds to 50\% left-trimmed average MDM values of each hexagon whilst the one on the bottom shows the top 10 highest pixel-level MDM values as scatter points
. 
Each hexagon has a width of 5 km and an area of $\approx$22km$^2$.}
\label{fig:hond}\vspace{-0.35cm}
\end{figure}

All these three aforementioned test locations are sharing the same basis in terms of marine pollution that historically, either no or low mass marine pollution problems reported in these areas, and all three test locations are going to be used to test the MAP-Mapper in no/low-density MD\&SP coverage. 
We summarise the MAP-Mapper findings for these test locations as:
\begin{itemize}
    \item Density maps of all three locations presented in Figure \ref{fig:hond} show similar characteristics as expected. Nearly all the hexagonal regions have highly bright colour coding that corresponds to no/low MD\&SP density with MDM values less than 0.5. 
    \item Chubut highest pixel MDM values are also very low and close to 0.00 whilst Hong Kong has some values around 1.50 MDM. 
    \item Even though it, in general, shows low MDM and MD\&SP density mapping findings, the Cornish coastline data has some high pixel-level MDM detections ($\approx$7.00), especially near Falmouth estuary (10 in Figure \ref{fig:hond}), that appears to coincide with large quantities of boats at their moorings.
\end{itemize}

\section*{Discussion}
Previous work in the marine debris detection research area has demonstrated that the detection of MD\&SP using satellite data and machine learning is possible. However, the capabilities of these approaches have not yet been explored on how they can be applied to real-world data for the purpose of automated mapping of marine debris density. With the development of MARIDA data set\cite{kikaki2022marida}, it has now been possible to train new machine learning models and evaluate their effectiveness in density mapping. Consequently, we developed an automated tool - \textit{MAP-Mapper} - to assess suspected plastic concentration and to highlight patterns of debris distribution,
and identify key areas of aggregation.

Both \textit{-Opt} and \textit{-HP} versions of the MAP-Mapper obtained considerable results and performance improvements compared to the MARIDA baseline models as seen in Table \ref{tab:res1}. In particular, MAP-Mapper models improved the MD\&SP detection performance up to 16\% and 5\% in terms of the precision and $F_1$-Score metrics compared to the Kikaki et. al. baseline models. On the other hand, the validation of the Plastic Litter Project targets has been highly promising. These results show that the model can detect plastic of different types. Furthermore, at lower thresholds, the model is able to detect plastic on the sub-pixel scale. It is unsurprising that very few plastic pixels were detected at a threshold of 0.99. Man-made targets often only covered a small percentage of a pixel. For the larger target, much higher probabilities were assigned to pixels containing 100\% plastic. Additionally, the targets were made of a single type of plastic, and contained no other types of debris. In the marine environment, plastic patches are extremely unlikely to be this uniform. They usually consist of many different types of plastic and/ or a mix of natural organic material\cite{biermann2020finding,ciappa2022marine}. For this reason, the training data is likely to follow this pattern and thus, spectral characteristics of the training data and plastic targets differ to some degree. Finally, the shape and size of the targets are significantly different from the plastic patches in the MARIDA dataset. These contextual factors may help explain why many of the plastic pixels are not predicted as plastic when using high thresholds. Furthermore, our analysis demonstrates that F-mask has a tendency to mask sargassum patches. Presumably, this is because its spectral signature is similar to terrestrial vegetation, and this is masked as land. Plastic is commonly found embedded within sargassum patches and therefore it is unsurprising that both plastic and sargassum can be found in the same feature\cite{ciappa2022marine}. Analysing the prediction files before and after masking found that F-mask did not mask any pixel that was identified as suspected plastic. This is a promising finding which suggests that the F-mask is surely a suitable algorithm for plastic detection by not masking suspected plastic pixels.

\begin{figure}[ht!]
\begin{minipage}{\textwidth}
    \centering
    \subfigure[]{\includegraphics[width=0.35\linewidth]{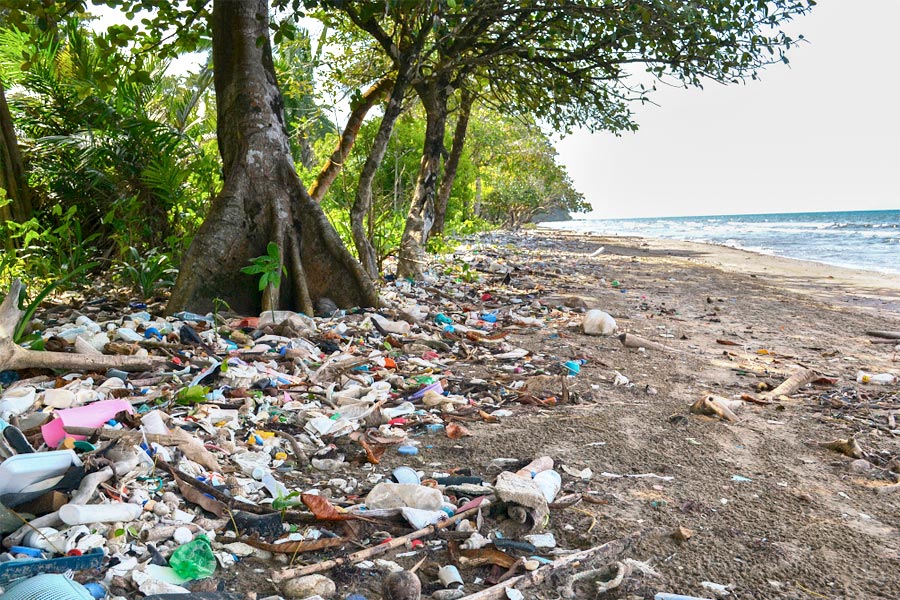}}
    \subfigure[]{\includegraphics[width=0.35\linewidth]{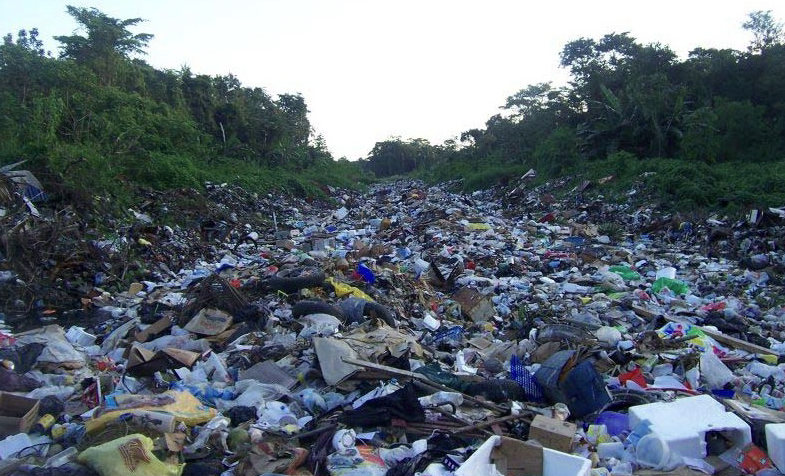}}
    \subfigure[]{\includegraphics[width=0.35\linewidth]{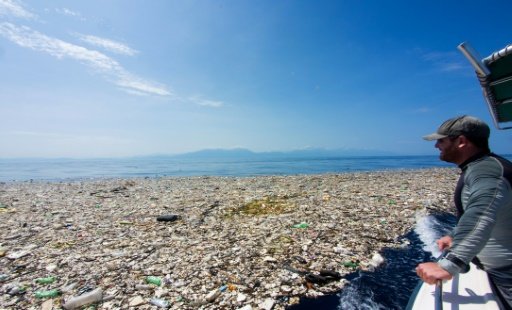}}
    \subfigure[]{\includegraphics[width=0.35\linewidth]{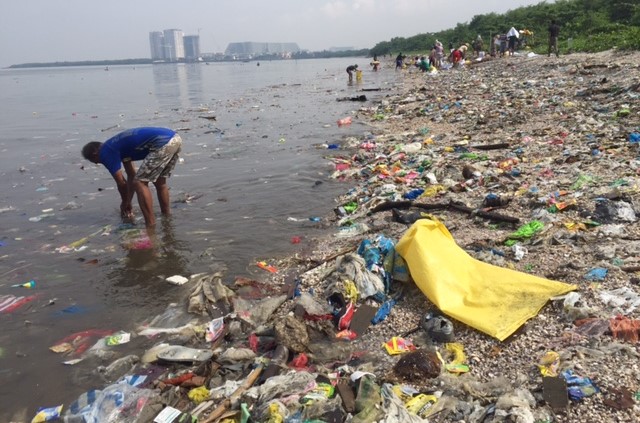}}
    \subfigure[]{\includegraphics[width=0.35\linewidth]{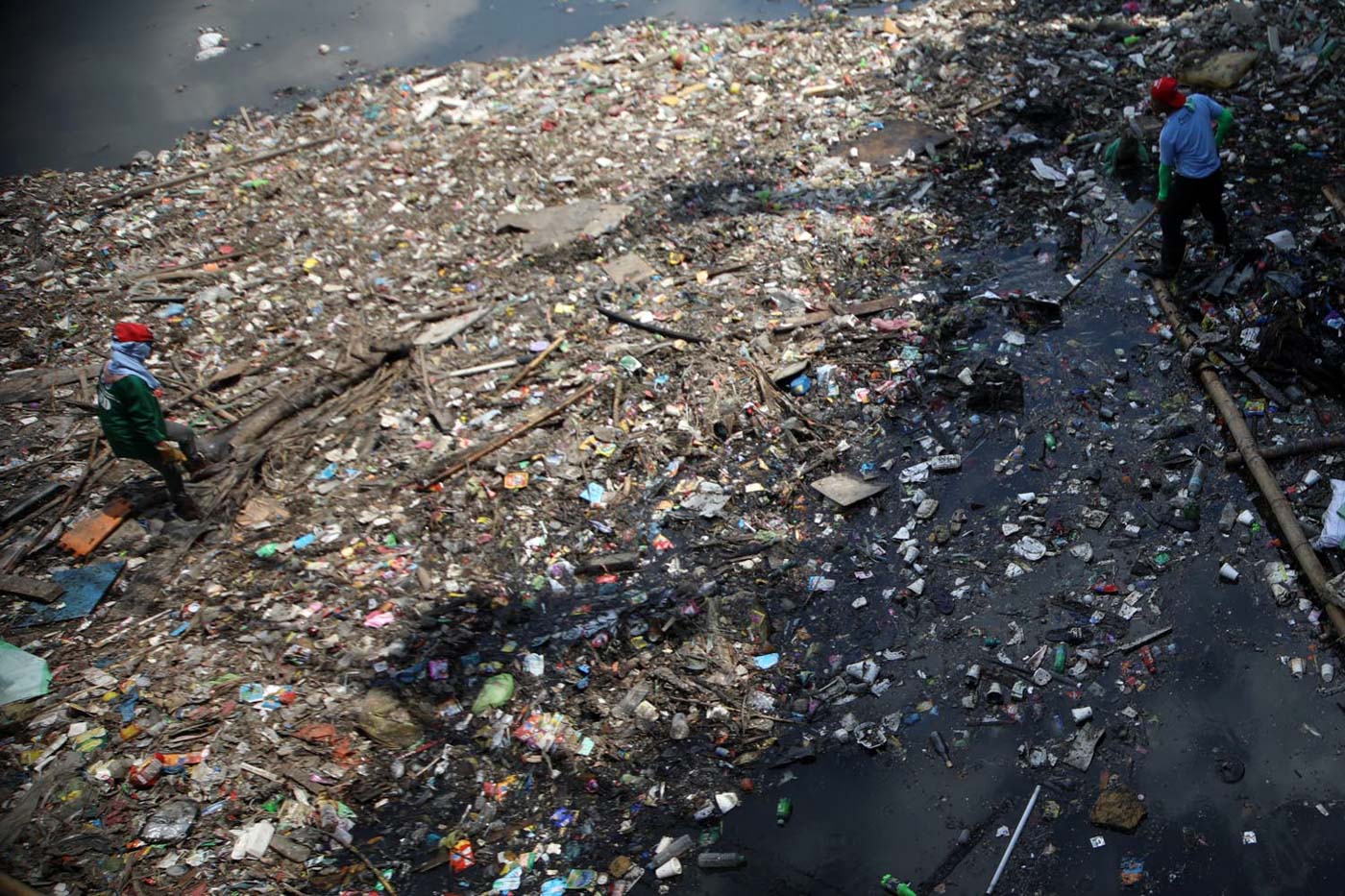}}
    \subfigure[]{\includegraphics[width=0.35\linewidth]{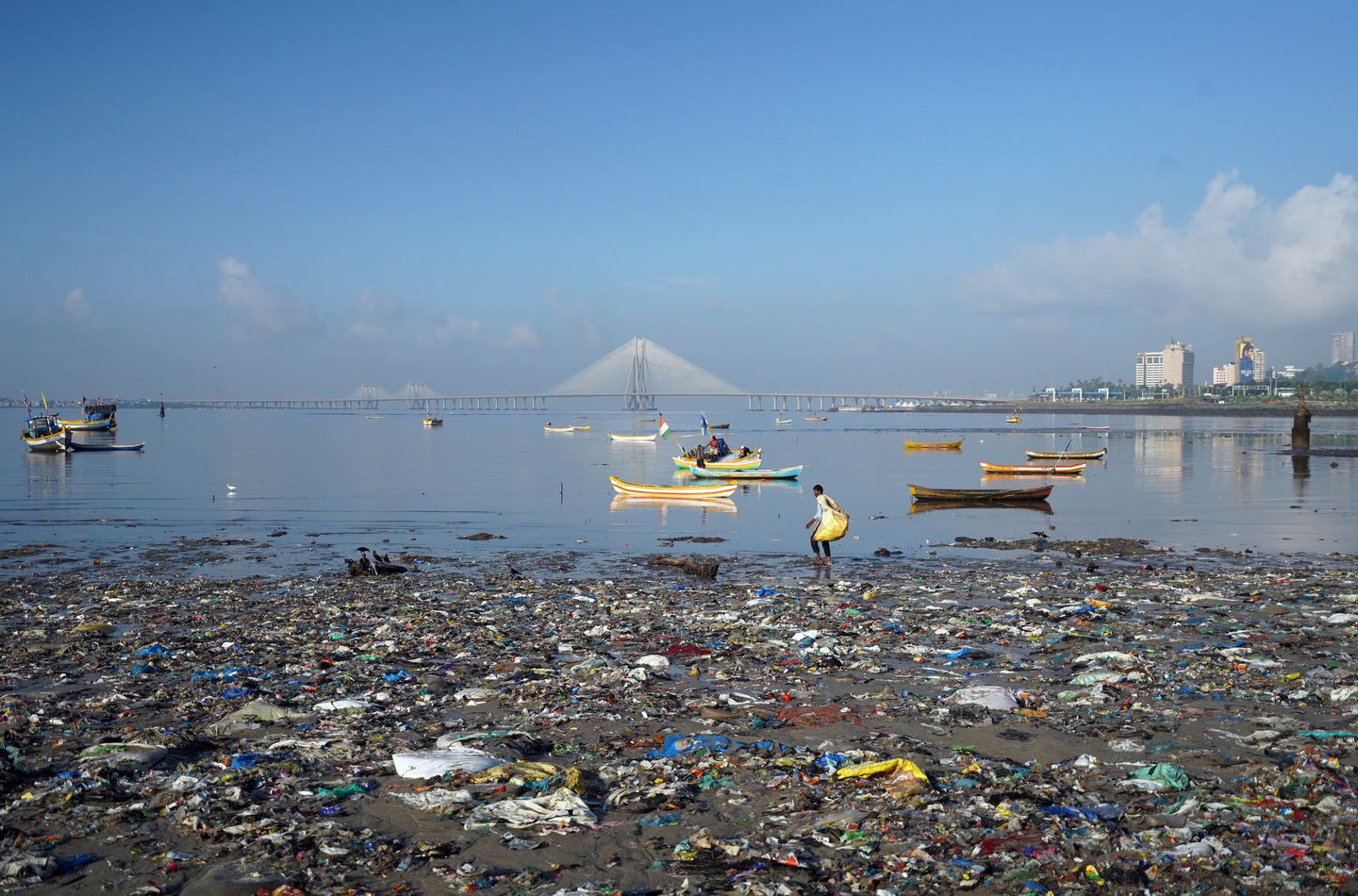}}
    \subfigure[]{\includegraphics[width=0.35\linewidth]{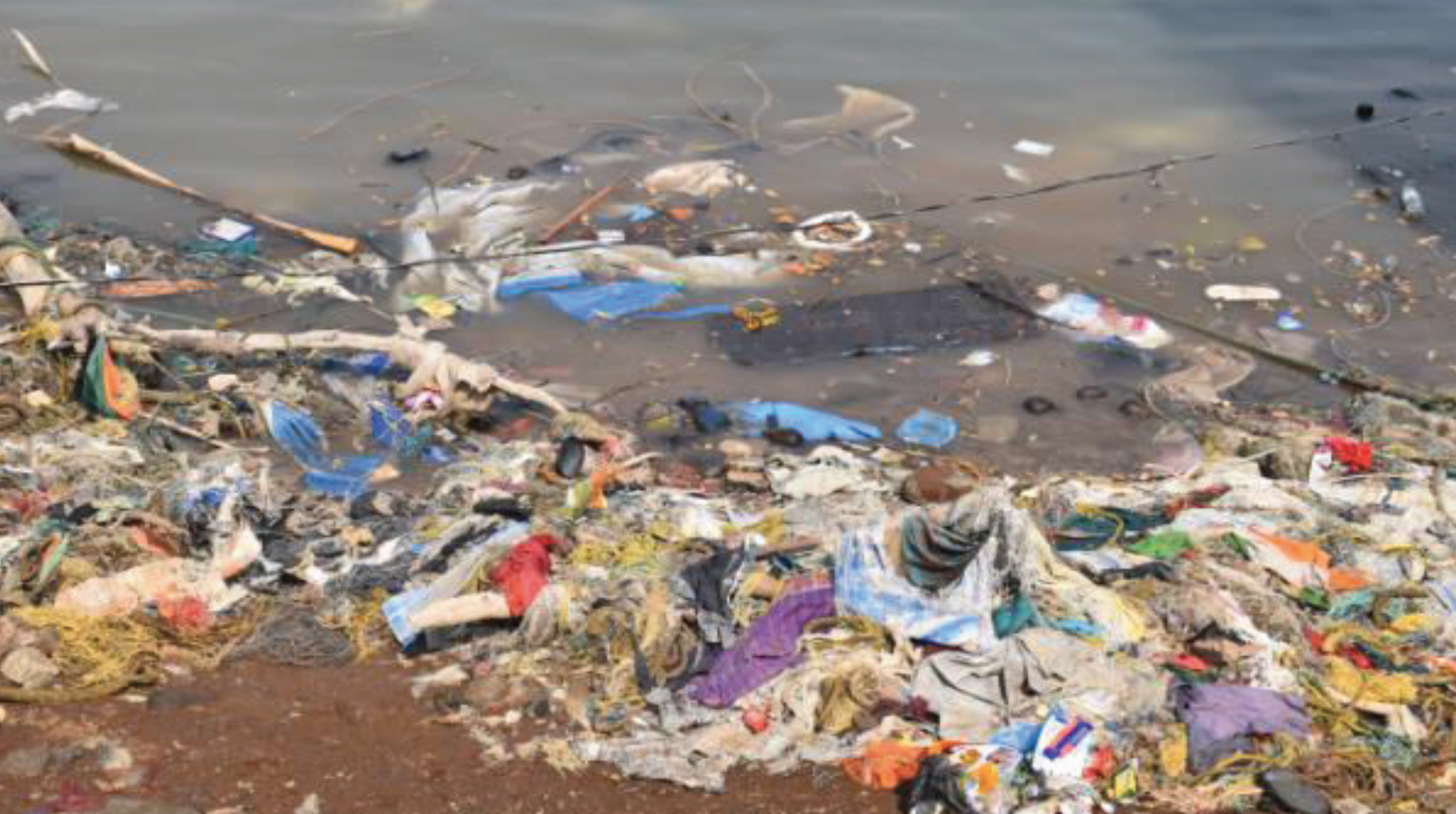}}
    \subfigure[]{\includegraphics[width=0.35\linewidth]{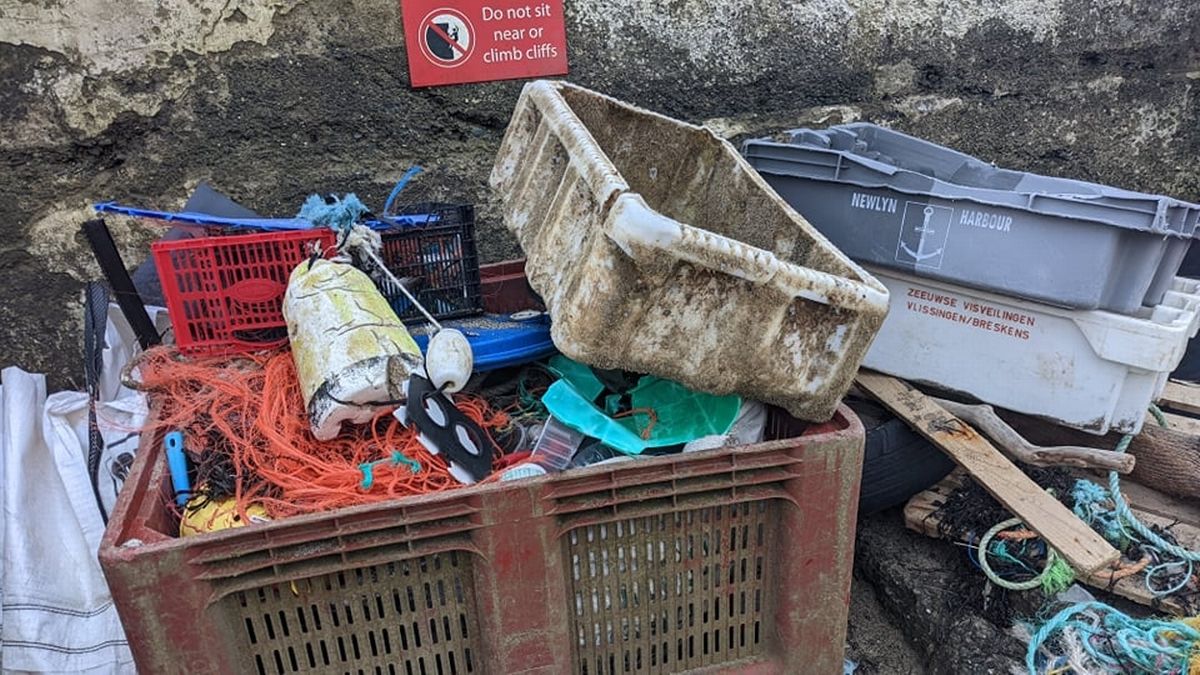}}
    \caption[]{Example real photographs for high-density detections of MAP-Mapper models. (a) Livingston, Bay of Amatigue \cite{jazzy_2022}, (b) Punta Gora, Bay of Amatigue \cite{lu2013ocean}, (c) Omoa, Gulf of Honduras\cite{leiva_2017}, (d) Freedom Island, Manila\cite{plasticmap_2022}, (e) Pasig River, Manila\cite{afp_2018}, (f) Mahim Bay, Mumbai\cite{editorialboard_2019}, (g) Panvel Creek, Mumbai \cite{naik2021assessment}, (h) Lizard Peninsula, Cornwall, UK\cite{becquart_2022}.}
    \label{fig:real}
  \end{minipage}\vspace{-0.35cm}
\end{figure}

Despite this promising performance of the proposed approaches, manual inspection is needed to be conducted to investigate areas of high plastic density and verify the results of the out-of-distribution test locations where possible. This would help ascertain the strengths and weaknesses of the proposed models. The Bay of Amatique (1 in Figure \ref{fig:hond}) in the Gulf of Honduras was found to have a high density of MD\&SP in two different areas as shown in Figure \ref{fig:hond}. The coast off Macho Creek between Livingston and Puerto Barrios is one of the potentially MD\&SP gathering points for the litter coming out of Rio Dulce that is the source of 16 million kg of mismanaged plastic per year \cite{cleanup_2022}. The Bay of Amatique has an offshore gyre that rotates counterclockwise and potentially carries the Rio Dulce and other sources of debris to the bays of Puerto Barrios and Bahia La Graciosa. Apart from the Bay of Amatique, the coasts of Omoa (3 in Figure \ref{fig:hond}) and Punta Gorda (2 in Figure \ref{fig:hond}) near Maya Mountains Marine Corridor appeared as the highly densely polluted areas which follow academic research for these areas \cite{lu2013ocean} (See Figure \ref{fig:real}). With being home to 4 of the world's 6 most polluted rivers, Manila of the Philippines has become the highest densely MD\&SP polluted place in MAP-Mapper analysis. Cavite City near the Freedom Islands (6 in Figure \ref{fig:hond}) has the highest average MDM in the analysis one of the reasons for which is the Paranaque River. This river and its branches are responsible for 10 million of kg mismanaged marine plastic pollutants \cite{cleanup_2022}. MAP-Mapper also highlighted the mouths of the world's first two highest-polluted rivers of Pasig and Tullahan (5 in Figure \ref{fig:hond}). These two rivers are responsible for 550 and 96 million mismanaged marine plastic pollutants, respectively \cite{cleanup_2022}. Similarly to the Gulf of Honduras and Manila results, the proposed MAP-Mapper density mapping tools achieved realistic detection results in Mumbai, India test location. Mahim Bay (7 in Figure \ref{fig:hond}) with one of the highest average MDM-valued places in the analysis, also known to be one of the most polluted bays in the world. The world's third plastic-polluted river Ulhas is also located in Mumbai, and the mouth of this river which is opened to Thane and Panvel Creek (8 in Figure \ref{fig:hond}) is also mapped as a highly polluted area by the MAP-Mapper tools with MDM values higher than 2.00. 

All the aforementioned explanation shows us that the developed MAP-Mapper tools are highly realistic and generalisable to out-of-distribution data. The results also suggest that Map-Mapper Tool is clearly useful for mapping MD\&SP density in areas of high pollution. In these regions, the presence of false positives is far less detrimental to the overall density map. True positives appear to outnumber false positives in each of the three highly polluted test locations above, thus suggesting that MAP-Mapper is a useful tool for mapping the MD\&SP density regions like this. In order to show MAP-Mapper's applicability in regions with known lower plastic pollution, we investigated Cornish Coastline, UK, Hong Kong, and Chubut, Argentina. The results presented in the previous section for these regions show that MAP-Mapper maps these regions with highly low mean MDM values. Considering the regions of interest for each test location historically do not have important mass MD\&SP pollution events, this is parallel with our results and provides shreds of evidence for the global applicability of the MAP-Mapper tools. Cornish Coastline somewhat diverges from Hong Kong and Chubut in terms of pixel-level MDM detections. Both of these regions have the highest pixel level MDM values of around 1.00 whilst Cornish Coastline has values higher than 5.00. The detailed analysis in this region suggests that these pixels are most likely waves and/or sea foam. These features follow the linear trajectory that is typical of the marine debris pixels in the MARIDA dataset but the prevalence leads to the conclusion that some of these detections are likely to be false positives. Although it is not possible to verify every pixel, it is likely that the MARIDA dataset does not contain enough examples of sea foam (0.15\% of the whole data set) for the model to differentiate effectively between foam/waves and plastic in this region. Another clear example of miss-classification is found in the Falmouth estuaries (10 in Figure \ref{fig:hond}). Some of the small boats are potentially misclassified as MD\&SP. Interestingly, larger boats with wakes were not misclassified. This suggests that more training data is needed to improve model performance when differentiating MD\&SP from some types of static watercraft. The results for Cornish Coastlines, especially for pixel level detections around the Lizard (11 in Figure \ref{fig:hond}), St. Ives \& St. Agnes (12 in Figure \ref{fig:hond}), can also be read as evidence to the work by Nelms et. al. \cite{nelms2020investigating} stating that the beaches of these regions are potentially the highest litter accommodating beaches in England. 

Figure \ref{fig:real} and Table \ref{tab:real} show eight example regions that MAP-Mapper models recognised as high MD\&SP density problems with their corresponding average and highest MDM values. Figure \ref{fig:real} shows real photographs of MD\&SP problems in each of these ROIs, and provides evidence regarding the generalisation capabilities of the proposed MAP-Mapper models. The importance of these representations is that, even though the test data is coming from out-of-distribution locations, highly valued MDM density map locations are aligned with the real field studies. 

\begin{table}[ht!]
    \centering
    \begin{tabular}{p{3.5cm}p{6.5cm}p{1cm}p{1cm}p{1.5cm}p{1.5cm}}
    \toprule
      Test Location &		Location details & Avg MDM	&Highest MDM&	Position in Fig. \ref{fig:hond} &	Evidence\\\toprule
      	& A beach near Livingston, Guatemala & 1.87 & 3.83 & Area 1 & Fig. \ref{fig:real}-(a)\\\cline{2-6}
      Gulf of Honduras&Punta Gora - Maya Mountain Marine Corridor&	1.51	&3.18	&Area 2	&Fig. \ref{fig:real}-(b)\\\cline{2-6}
	&Off the coasts of Omoa	&0.96	&3.21	&Area 3	&Fig. \ref{fig:real}-(c)\\\hline
Manila, Philippines&	Coasts of Freedom Island	&2.90	&9.15	&Area 6	&Fig. \ref{fig:real}-(d)\\\cline{2-6}
	& Pasig River	&1.90	&16.29	&Area 5	&Fig. \ref{fig:real}-(e)\\\hline
Mumbai, India&	Mahim Bay&	2.40	&9.43	&Area 7	&Fig. \ref{fig:real}-(f)\\\cline{2-6}
	&Panvel Creek	&2.07	&8.89	&Area 8	&Fig. \ref{fig:real}-(g)\\\hline
 Cornish Coastline, UK & Polurrian beach on the Lizard Peninsula & 0.43 & 6.99 & Area 11 & Fig. \ref{fig:real}-(h)\\
\bottomrule
    \end{tabular}
    \caption{Some regions recognised as high MD\&SP density by the MAP-Mapper models.}
    \label{tab:real}
\end{table}

MAP-Mapper is an initial product from a detailed remote sensing computational imaging project for the purpose of detecting and tracking MD\&SP. Our next steps to improve on MAP-Mapper Tool(s) would include (non-exhaustive):

(1) In order to improve the global applicability of MAP-Mapper and to address issues with potential miss-classifications, we plan an expansion of the MARIDA dataset particularly from under-represented regions, and predominantly focus on the inclusion of more plastic, foam, ships/boats, and cloud pixels, as well as pixels affected by sun glint. 
    
(2) MAP-Mapper was designed to provide insights into the concentration, distribution, and pathways of marine plastic. Therefore, to help enhance scientific analysis, historical weather queries will also be integrated into the analysis stage. This would make it possible to investigate how recent rainfall correlates with riverine plastic output, especially in locations with poor waste management, e.g. around the mouths of River Motagua in Guatemala and River Pasig in Manila.
    
(3) MAP-Mapper would benefit from a better assessment of its validity for MD\&SP density mapping. To achieve this, suspected plastic detections should be explored with very high-resolution satellite data, ground truth reports, or aerial imagery. This would help certify that pixels are not being miss-classified and help to determine their validity.
    
(4) The model trained with only 4 bands (the ones contributing to calculating FDI and NDVI) demonstrated good performance than using all 13 Sentinel-2 bands. However, it is possible that other band combinations could improve model performance. Removing some of the lower resolution bands, as well as bands where wavelengths do not correlate with plastic materials, may reduce noise in the data set. Development of future models should trial different band combinations to assess which bands provide the best performance, alongside run-time reduction. 
    
(5) As suggested by Chuanmin Hu\cite{hu2022remote} in their cautious note on spectral interpretations, spectrally distorted shapes have an important impact on the visibility of suspected plastics from multi-spectral satellite imagery. For this, we plan to conduct an investigation on the validity of satellite imagery spectral returns with in-situ real marine debris and floating plastics targets.

(6) From a technical point of view, similar to MARIDA baseline models, the MAP-Mapper architectures utilise supervised learning approaches. These are highly dependent on the labelled training data, and any problem in the training data set, such as low confidence labels or highly unbalanced class distribution might cause wrong classifications. In the following versions of MAP-Mapper tools, we promote using minimal-supervision (semi-, self- and/or un-supervised) approaches that exploit the usage of unlabelled data to extract further useful information. 

\section*{Methods}
\subsection*{Measuring Density with a Marine Debris Metric} 
Developing a density map only taking into consideration the number of detected MD\&SP pixels is not suitable considering pixel probabilities in single-date imagery might be high due to several reasons: e.g. misclassifications, dynamic nature of MD\&SP, etc. A combination of this with the long time interval and the high number of images for each ROI will strongly affect  the accuracy of the MD\&SP density maps and create highly wrong meanings. In order to solve this problem - in the context of density mapping in a time interval - we develop a metric that promotes average pixel probabilities and detection threshold values:
\begin{align}
    MDM_{ij} &= \left[\dfrac{100}{N}\sum_{k=1}^N \mathbbm{1}_{p_{ijk}\geq T}\right] \times
    \left[\dfrac{1}{N}\sum_{k=1}^N p_{ijk}\right]\\
    &= \mathcal{D}_{ij}\times \bar{\mathcal{P}_{ij}}
\end{align}
where $\mathbbm{1}_{p_{ijk}\geq T}$ is the indicator function that returns 1 when the condition $p_{ijk}\geq T$ is satisfied, and 0 otherwise. The pixel-level probability of being MD\&SP is denoted as $p_{ijk}$ on pixel ($i,j$) and date $k$ whilst $T$ refers to the threshold value to determine whether a pixel is MD\&SP or not. Thus, $\mathcal{D}_{ij}$ becomes the percentage of the number of detections within the selected time interval whilst similarly $\bar{\mathcal{P}_{ij}}$ refers to the average probability of being MD\&SP on pixel ($i,j$) for the given time frame.

From the above description, we can see that $MDM_{ij}$ is a positively valued metric where 0 means no MD\&SP problem, and a higher $MDM$ value corresponds to a polluted location on the maps. On the other hand, $MDM$ can be seen as a metric in which the average probability value of a pixel is positively weighted if the total number of detections is high in that corresponding pixel (e.g. 10 dates out of 20 dates in the time interval means 50\%). This is giving us a better picture to measure the MD\&SP density maps in a global time scale with the MAP-Mapper approach. $MDM$ can also be seen to be a universal metric to compare different locations on the earth considering the utilisation of the fixed area of hexagons and the averaging over the given time interval.
    
\subsection*{MAP-Mapper Pipeline}
\begin{figure}[ht]
\centering
\includegraphics[width=0.8\linewidth]{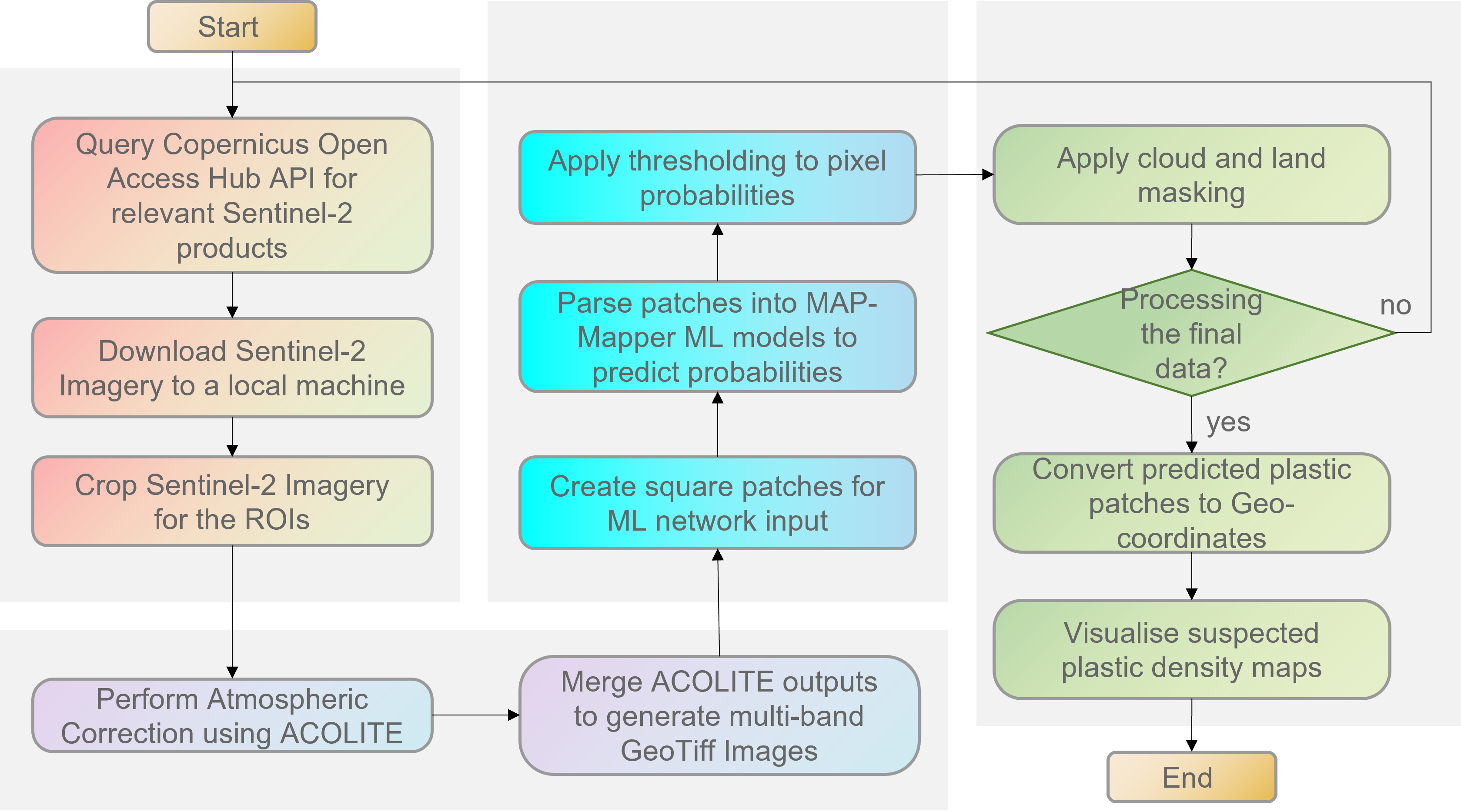}
\caption{Flow diagram showing the MAP-Mapper pipeline. The above system is fully automated via the Python terminal entering a single command line including the ROI coordinates and several parameters of the network. }
\label{fig:flow}\vspace{-0.35cm}
\end{figure}

The development of the tool was divided into separate components. These components were then integrated and conducted in sequence to produce outputs for the ROI. A flow diagram of this process can be seen in Figure \ref{fig:flow}.

\textbf{Data gathering:} Since the MARIDA dataset consists of Sentinel-2 data, MAP-Mapper requires first to collect Sentinel-2 data for an ROI and date range. This is achieved by querying the Copernicus Open Access Hub API for the required Sentinel-2 data. Querying requires entering corner coordinates of the selected rectangle ROI and desired time interval. MAP-Mapper then downloads Sentinel-2 imagery to a local machine for further processing.

\textbf{Atmospheric correction:} Once the data is downloaded, the ACOLITE software is used to perform the atmospheric correction. This ensured that both training data and input data were corrected with the same algorithm. The ACOLITE is a suitable tool for atmospheric correction of coastal and ocean regions and its validity for use in marine-plastic detection has been previously investigated and verified\cite{topouzelis2019detection,kikaki2020remotely}. Dark spectrum filtering is used to ensure consistency with the training data set and sun glint correction is also applied.
    

\textbf{Machine learning step for prediction and thresholding:} Following the atmospheric correction, the ACOLITE outputs are combined into one large multi-banded GeoTiff file and input patches are created. These patches are then given to the MAP-Mapper network architectures to make predictions showing a probability of each pixel in an image being MD\&SP. Therefore, thresholding is applied to produce binary classification outputs of “MD\&SP” or “not MD\&SP”. 


\textbf{Cloud and Land Masking:} It has been previously shown that cloud masking reduces not only the presence of clouds in analysis but also the chance of miss-classification \cite{qiu2019fmask}. For the purpose of MAP-Mapper density mapping, water pixels are kept and all other pixels are then masked. However, this does not provide robust land masking. Therefore, land masking is also implemented by cropping a worldwide geospatial vector file to the ROI. The output of the abovementioned process is then used to mask all land so that only ocean regions are kept for further analysis. 


\textbf{MDM Calculation and Density map creation:} In order to obtain MD\&SP density maps, we first obtain probability (direct network output) and detection (thresholded to become a binary output) maps. Each pixel coordinate is then converted to coordinate reference system points, consisting of a longitude and latitude where their corresponding probability and detection values are used to calculate a novel metric of \textit{MDM} to visualise the MD\&SP density. 


\textbf{MD\&SP Density Map Visualisations:} For each pixel coordinate, the calculated MDM values are then plotted on the map using hexagonal binning. Regardless of their area, for each test location, the width of a hexagon is set to 5km and for all plotted images each hexagonal area was fixed at approximately 22 km$^2$. In order to better quantify the density mapping findings 50\% left trimmed average MDM values are used for each hexagon whilst colour-coding the findings. 

\subsection*{MAP-Mapper Network Architecture}
Starting from a generic U-net architecture, we performed several optimisation/improvement stages in order to develop the MAP-Mapper architectures such as changes in input channels, output channels, batch size, patch size, thresholding, and the number of feature maps in each layer.

The first novel improvement has been achieved by reducing the utilisation of all Sentinel-2 bands. Biermann et. al.'s FDI\cite{biermann2020finding} and NDVI indexes only require 4 of the spectral bands available in Sentinel-2 images. This suggests that some of the bands encode more salient information about material type than others. Also, other bands may be introducing noise as they may not correlate with material type. Thus, extra bands may increase computational complexity, without improving detection. For these reasons, MAP-Mapper machine learning architectures are trained with only 4 input Sentinel-2 bands of 4, 6, 8, and 11.  In our initial tests, this provided promising results and was just as effective as a model trained with all bands. This significantly reduces (i) the amount of data required for MD\&SP detection, and (ii) MAP-Mapper architectures' run-time.

In architectures like U-net, feature maps of the first layer extract simple features whilst the feature maps in later levels extract increasingly complex features. Consequently, even though increasing the number of feature maps increases the number of model parameters, this typically increases the model's ability to learn. On the other hand, complex models have a tendency to over-fit, take longer to train, and take longer to make predictions on input data. Therefore, a balance is required. The MAP-Mapper development stages showed us the highest number of feature maps among 64, 128, 256, and 512 for each corresponding encoder layer achieved the best performance, so thus set as the MAP-Mapper architecture parameter.

We found out that (parallel with the expectations) smaller batch and patch sizes were correlated with significantly more gradual and longer model training times. It was found that smaller patches and batch sizes resulted in better model performance. Therefore, MAP-Mapper architectures consist of a batch size of 64 and a window size of 32.

Binary classification requires a threshold to discriminate whether the class is MD\&SP or not. In general, the decision boundary line was set at 0.5. Therefore, any pixel that is predicted to be MD\&SP with a probability greater than 0.5 was classified as MD\&SP. Thresholding changes this value to influence the recall and precision of a model. A precision-recall curve was created to enable the identification of the optimal threshold, which resulted in the threshold of 0.815 producing an $F_1$ of 88\% with a precision of 0.87 and a recall of 0.88 (Threshold of the MAP-Mapper-Opt architecture).


Map-Mapper-Opt precision value of 0.87 can be seen as low for a reliable MD\&SP monitoring application. This is because of concerns about false positives being detrimental to the validity of the results. Therefore, the threshold was increased further to achieve higher precision values. MAP-Mapper-HP achieves a precision of 0.95 with a threshold of 0.99 whilst still maintaining a reasonable recall value of 0.63 ($F_1$ of 75\%).


\bibliography{MAP-Mapper-Refs}



\section*{Acknowledgements}
The authors want to thank Robin de Vries, from The Ocean Cleanup, for taking an interest in our work and giving us helpful advice on this challenging topic. The authors would also like to thank Prof Chuanmin Hu for their valuable suggestions about spectral characteristics of marine debris. This work was performed using the computational facilities of the Advanced Research Computing at Cardiff (ARCCA) Division, Cardiff University.

\section*{Author contributions statement}

H.B. and O.K. wrote the main manuscript text, W.M. created tables, H.B. created the data pipeline and the baseline code, H.B. conducted evaluation and validation experiment(s), W.M. and H.B. conducted test location experiments, O.K. developed the MDM metric, H.B., W.M. and O.K. created data visualisation outputs, analysed the results, and reviewed the manuscript. O.K. supervised the research.

\section*{Competing interests}
The author(s) declare no competing interests.

\section*{Additional information}
The Python code for the MAP-Mapper software will soon be published on the authors' GitHub pages. 

\noindent \textbf{Correspondence} and requests for materials should be addressed to O.K.




\end{document}